\documentclass[prf,longbibliography]{revtex4-2}
\usepackage[
paperwidth=178mm, 
paperheight=254mm, 
bottom=15mm,
top=25mm,
left=20mm,
right=20mm]{geometry}
\usepackage{graphicx}
\usepackage{epstopdf, epsfig}
\usepackage{hyperref}
\usepackage[usenames,dvipsnames,svgnames,table]{xcolor}
\usepackage{mathptmx} 
\usepackage[utf8]{inputenc}
\usepackage{newtxmath}
\usepackage{natbib}
\usepackage{microtype}
 
\hypersetup{
colorlinks=true,
linkcolor=blue,
urlcolor=blue,
filecolor=blue,
citecolor=blue
}

\newcommand{\vect}[1]{\boldsymbol{#1}}

\newcommand\Ri{\mbox{\textit{Ri}}}
\newcommand\Rey{\mbox{\textit{Re}}}
\newcommand\Sc{\mbox{\textit{Sc}}}

\begin{document}
\title{Two regimes of dilute turbulent settling suspensions under shear}

\author{Jake Langham}
\email[]{j.langham@bristol.ac.uk}
\author{Andrew J. Hogg}
\affiliation{School of Mathematics, Fry Building, University of Bristol,
Bristol, BS8 1UG, UK
}

\date{\today}

\begin{abstract}
When turbulent flow is laden with negatively buoyant particles, their mean
distribution in the direction of gravity can induce stable density gradients
that penalize turbulent fluctuations. This effect is studied numerically for
shear-driven flow with dilute non-inertial sediment. The turbulent dynamics and
sediment transport depend critically on particle settling velocity $v_s$, splitting
into two regimes: homogeneous weakly stratified turbulence and flow with
developed turbulence atop an intermittent boundary layer. At intermediate $v_s$,
neither state can be sustained and the flow laminarizes.
\end{abstract}

\maketitle

It is common for flows in nature to carry suspensions of fine particles over
great distances. Examples abound in the atmosphere, rivers, coastlines and on
the ocean
floor~\cite{Schepanski2018,Korup2012,Martinelli1989,Rijn2013,Mulder2003,Meiburg2010}.
Transport fluxes of suspended sediments have long been
studied~\cite{Rijn1984,Dyer1988,Soulsby1997}, due to their practical importance
in engineering, as well as the substantial contribution they make to the Earth's
sediment cycle~\cite{Elfrink2002,Gislason2006,Dethier2022,Syvitski2022}.  It is
the turbulent fluctuations within a flow that are responsible for lifting
particles into suspension and resisting their downward settling under gravity so
that they can be transported. Less widely appreciated are the effects of
suspensions on turbulence itself. Since work must be done by the flow 
against
gravity in order to keep negatively buoyant sediment in suspension, even very
dilute concentrations of settling particles are reported to reduce the intensity
of turbulent
fluctuations~\cite{Villaret1991,Winterwerp2001,Tu2019,Egan2020,Huang2022}.

Efforts to probe this effect in detail using direct numerical simulations (DNS) of
turbulence are far less common than for the related case of flows stratified by
gradients in temperature or solute concentration, where the onset, development
and characteristics of turbulence have been extensively studied for different
canonical flow configurations (see
e.g.~\cite{Caulfield2000,Smyth2000,Garcia2011,Brethouwer2012,Deusebio2015,Zhou2017,Lucas2019,Kaminski2019}).
For dilute sediment suspensions, research in this direction has predominantly
focussed on channel flow models of gravity-driven turbidity
currents~\cite{Cantero2009a,Cantero2012a,Cantero2012b,Shringarpure2012,Ozdemir2018}.
Investigation of turbulence suppression reveals the essential mechanisms at
play: clouds of settling particles preferentially concentrate towards lower
depths and extract turbulent kinetic energy from the flow via the vertical
fluxes required to maintain their average elevation. Suspensions with settling
velocities or bulk densities exceeding a critical threshold laminarize the flow due
to the coupled contributions of stable density stratification and localization
of the flow driving force towards the bottom
wall~\cite{Cantero2012a,Shringarpure2012}.
Simulations of pressure driven~\cite{Cantero2009b,Dutta2014} and oscillatory
channels~\cite{Ozdemir2010}, where this latter effect is not present, have
hinted at similar bounds for partial or full laminarization, contingent on the
suspension properties.
In this Letter, we report the case of shear-driven (plane Couette) flow. A new
turbulent regime is identified that exists beyond the theoretical laminarization
boundary and its properties are investigated. Flows in this regime exhibit a
boundary layer region containing most of the sediment, whose turbulent dynamics
are highly intermittent. This property is found to greatly alter statistically
steady sediment fluxes and their spatiotemporal distribution.

We consider incompressible fluid sheared between two infinite parallel planes
perpendicular to gravity---a stationary basal surface and an upper wall moving
with velocity $2U\vect{e}_x$.  The flow is laden with negatively buoyant
sediment particles, which settle under gravity with characteristic velocity $V_s
\vect{e}_y$. We assume the sediment to be sufficiently small and dilute that it
occupies a continuous phase within the channel, for which the physics of
particle inertia, cohesion and inter-particle collisions may be safely
neglected~\cite{Necker2002,Necker2005,Shringarpure2012}. Provided that density
variations within the mixture are small relative to the mean flow density,
the Boussinesq approximation applies and the system obeys the following
equations for the flow velocity $\vect{u}(\vect{x},t)$, pressure $p(\vect{x},t)$
and sediment concentration $c(\vect{x},t)$, rendered dimensionless with respect
to length and time scales $H$ and $H/U$, where $2H$ is the channel height, and
the fluid density:
\begin{gather}
    \frac{\partial \vect{u}}{\partial t} + \vect{u}\cdot\nabla{\vect{u}}
    = -\nabla p + \Rey^{-1} \nabla^2 \vect{u} - \Ri_b c
    \vect{e}_y,%
    \label{eq:governing 1}\\
    \nabla\cdot \vect{u} = 0,\\
    \frac{\partial c}{\partial t} + \vect{u}\cdot\nabla c - v_s
    \frac{\partial c}{\partial y} = \kappa \nabla^2 c.
    \label{eq:governing 3}
\end{gather}
The dimensionless parameters are the bulk Reynolds number $\Rey = UH / \nu$,
bulk Richardson number $\Ri_b = M (\varrho - 1) g H / U^2$, dimensionless
settling velocity $v_s=V_s/U$ and sediment diffusivity $\kappa = K / UH$, where
$\nu$ is the viscosity of the fluid, $g$ is gravitational acceleration, $M$ is
the mean volume fraction occupied by sediment, $\varrho$ is the ratio of
sediment and fluid densities and $K$ is an effective dimensional sediment
diffusivity which captures the aggregated effect of small, hydrodynamically
mediated fluctuations of individual particle
trajectories~\cite{Necker2002,Guazzelli2011,Shringarpure2012}.
Equation~\eqref{eq:governing 3} is normalized so that $c = \psi / M$,
where $\psi$ is the sediment volume fraction. For dilute suspensions, $\psi
\ll 1$.

We perform DNS of Eqs.~\eqref{eq:governing 1}--\eqref{eq:governing 3} within the
computational domain $[0, 4\pi]\times[0,2]\times[0, 2\pi]$.  Periodic boundary
conditions are enforced at $x = 0, 4\pi$ and $z = 0, 2\pi$. At the walls $y =
0,2$, we fix no-slip conditions for the fluid, $\vect{u}(x, 0, z) = \vect{0}$
and $\vect{u}(x, 2, z) = 2\vect{e}_x$, and the no-flux condition $v_s c + \kappa
\partial c / \partial y = 0$ for the sediment.  We use the pseudo-spectral
\texttt{Channelflow} code~\cite{Channelflow}, adapted to integrate
Eqs.~\eqref{eq:governing 1}--\eqref{eq:governing 3}, with $330 \times 256$
de-aliased Fourier modes in $(x,z)$ and $141$ Chebyshev modes in $y$.  The
parameters $\Rey = 3125$ and $\kappa = 3.2\times 10^{-4}$ are fixed throughout,
which implies Schmidt number $\Sc = (\Rey\kappa)^{-1} = 1$. This choice of
$\Sc$ follows estimates for sand particles in water adopted in turbidity current
models~\cite{Shringarpure2012}.  A smaller set of simulations were conducted in
the domain $[0,8\pi]\times[0,2]\times[0,4\pi]$ (increasing the in-plane
resolution to $660\times582$ modes) to confirm that the results presented herein
are insensitive to box size.

Throughout this Letter, quantities averaged over the horizontal coordinates $x$
\& $z$ are adorned with an overbar $\overline{\cdot}$, with primes denoting
fluctuations away from this mean, e.g.\ velocity fluctuations are $\vect{u}' =
\vect{u} - \overline{\vect{u}}$ and the total instantaneous turbulent kinetic
energy (TKE) for the flow is $\frac{1}{4}\int_0^2
\overline{\vect{u}'\cdot\vect{u}'} \,dy$.  Angular brackets $\langle \cdot
\rangle$ denote time averages, which are taken over at least 2000
advective time units when reporting DNS data.

To demonstrate the effect of the settling particle field on turbulence, we begin
with an illustrative numerical experiment. From a state of developed
unstratified turbulence, we increase $\Ri_b$ from $0$ to $0.06$, in increments
of $0.02$ separated by $200$ time units, for sediments with settling
velocities $v_s = 10^{-3}$, $2.5\times 10^{-3}$ and $4\times 10^{-3}$.
\begin{figure}
    \includegraphics[width=\columnwidth]{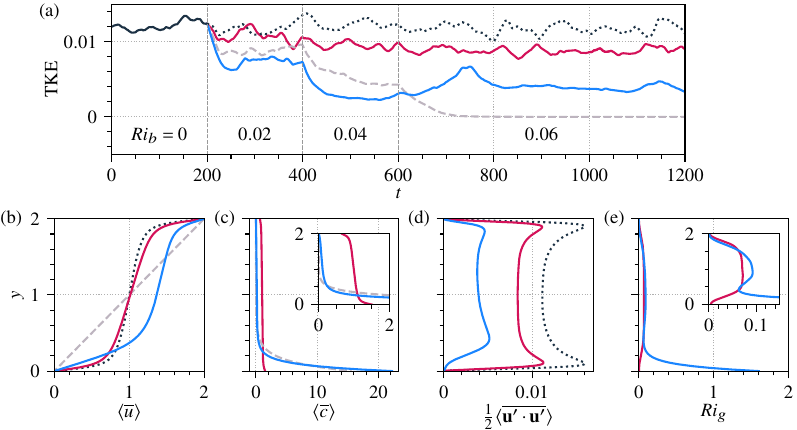}%
    \caption{%
    Development of turbulence into distinct regimes, depending on
    settling velocity, $v_s = 10^{-3}$ (solid red), $2.5 \times 10^{-3}$ (dashed
    gray) and $4\times 10^{-3}$ (solid blue). (a)~Total instantaneous TKE evolution for an
    initially unstratified flow (solid black, $0\leq t < 200$), incrementally
    subjected to increasing $\Ri_b = 0.02$ ($200 \leq t < 400$), $0.04$ ($400
    \leq t < 600$) and $0.06$ ($t \geq 600$). 
    For reference, the evolution of the $\Ri_b = 0$ case for $t \geq 200$ is
    plotted in dotted black.
    Panels~(b)--(e) show the
    wall-normal dependence of selected mean quantities, upon reaching a
    statistically steady
    state, at $\Ri_b = 0.06$:
    (b)~streamwise velocity,
    (c)~concentration, (d)~mean TKE and (e)~gradient Richardson number, $\Ri_g$.
    In panels (b,d), the corresponding statistics for $\Ri_b=0$ flow are included
    (dotted black).
    }
    \label{fig:profiles}%
\end{figure}
In Fig.~\ref{fig:profiles}(a), the resultant total TKE for the three flows is
plotted, alongside a reference case fixed at $\Ri_b = 0$ (dotted black).
The flows with lowest and highest $v_s$ reach statistically steady states,
following losses of TKE after the introduction of buoyancy effects.  In contrast, the
middle state cannot maintain turbulence and laminarizes.  The ultimate
steady-state streamwise velocity $\langle\overline{u}\rangle$ and concentration
$\langle\overline{c}\rangle$ profiles are shown in
Figs.~\ref{fig:profiles}(b,c).
Moreover, in Figs.~\ref{fig:profiles}(d,e), we show the $y$-dependence of both
the mean TKE and gradient Richardson number $\Ri_g = -\Ri_b
(d\langle\overline{c}\rangle/dy)/(d\langle\overline{u}\rangle/dy)^2$.
In the simulation with lowest $v_s$ ($10^{-3}$), the profiles exhibit
approximate symmetry about the centreline $y=1$.  Settling induces higher
concentrations towards the bottom wall, leading to an emergent bulk
stratification, which is reorganized by turbulence to leave profiles resembling
simulations of stably stratified shear flows~\cite{Deusebio2015,Zhou2017}.
Conversely, in the case of highest settling velocity ($4\times 10^{-3}$), most
of the sediment is contained within a narrow boundary layer ($0 \leq y \lesssim
0.4$).  Turbulent activity is suppressed over this region [see
Fig.~\ref{fig:profiles}(d)], as may be expected from the high concentration
gradients at the bottom wall.  Nevertheless, turbulence persists in the
relatively dilute upper channel, which feels a similar, but slightly stronger
level of stratification than the low $v_s$ case.  In the intermediate case,
$v_s$ ($2.5\times 10^{-3}$) is neither low enough that the flow becomes only
weakly stratified, nor high enough to drive sufficient quantities of sediment
out of the upper channel and turbulence is fully extinguished.

The turbulent flows described above are archetypes of two broad regimes we
identify from our simulations.  The first regime, which governs flows when $v_s$
or $\Ri_b$ are sufficiently small, we refer to as `weakly self-stratified' (WS)
and shares characteristics with corresponding DNS reported for pressure and
gravity-driven setups~\cite{Cantero2009a,Cantero2009b}.  The second regime,
which exists beyond the laminarization boundary for WS flows, is yet to be
identified in prior studies.  We refer to these flows, in which turbulence is
sustained above a strongly-stratified near-bed region, as sediment boundary
layer (SBL) turbulence. 

\begin{figure}
    \includegraphics[width=\columnwidth]{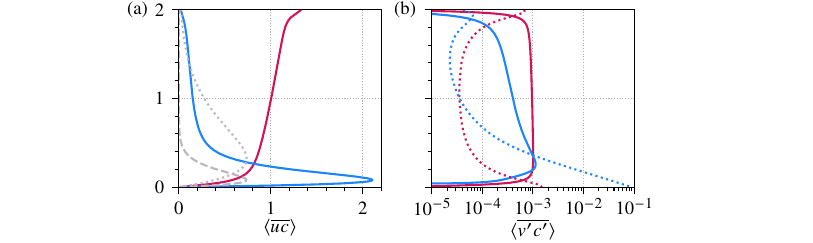}%
    \caption{Concentration correlations for  
    a WS flow with $v_s =
    10^{-3}$ (red) and SBL flow with $v_s = 4
    \times 10^{-3}$ (blue). In both cases, $\Ri_b = 0.06$.
    (a)~Streamwise sediment transport. The gray lines display the equivalent
    values for laminar flow with $v_s = 10^{-3}$ (dotted) and $v_s = 4 \times
    10^{-3}$ (dashed).
    (b)~Vertical turbulent sediment flux (solid lines). Also shown
    is $-\kappa d\langle \overline{c}\rangle / d y$
    (dotted lines).
    }
    \label{fig:transport}%
\end{figure}

Of particular interest in applications is the rate of sediment transported along
the channel.  In Fig.~\ref{fig:transport}(a), we plot the streamwise sediment
flux $\langle \overline{uc} \rangle$, for our example WS and SBL flows.
\begin{figure*}
    \includegraphics[width=0.98\columnwidth]{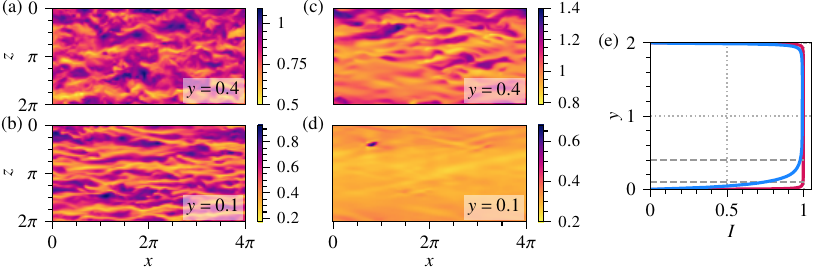}%
    \caption{Intermittency near the bottom wall.  (a)--(d)~Horizontal slices
    for the WS (a,b) and SBL (c,d) example flows,
    at $y = 0.4$ (a,c) and $y = 0.1$ (b,d),
    showing the instantaneous streamwise velocity field $u$ as a function of $x$
    and $z$. Snapshots (a,b) and (c,d) are taken at the same instant. 
    Videos showing the full simulations from which the snapshots in
    (a)--(d) were taken are available in the supplementary material.
    (e)~Dependence of intermittency factor $I$ on $y$ for the example WS 
    (red) and SBL (blue) flows. Dashed gray lines are at $y=0.1,0.4$.
    }
    \label{fig:intermittency}%
\end{figure*}
The flux for the WS flow monotonically increases with~$y$,
while the SBL flow transports most of its
sediment within the boundary layer at the bottom wall. Though the no-slip
boundary causes $u$ to be relatively low in this region, high near-wall
concentration leads to a peak streamwise flux at $y \approx 0.08$.  In both
cases, we find that the contribution from the turbulent streamwise flux
$\langle\overline{u'c'}\rangle$ (not plotted) is negligible
($|\langle\overline{u'c'}\rangle| < 0.025\langle \overline{uc}\rangle$).
Nevertheless, fluctuations are important due to the role they play in
adjusting the mean flow and concentration profiles away from laminar flow.  On
averaging the $x$-directed component of Eqs.~\eqref{eq:governing 1}
and~\eqref{eq:governing 3} over $x$, $z$ and $t$, one obtains:
\begin{gather}
    \frac{d\langle \overline{u} \rangle}{dy}
    -
    \frac{d\langle \overline{u} \rangle}{dy}
    \bigg|_{y=0}
    =
    \Rey \langle \overline{u'v'} \rangle,
    \label{eq:u mean}\\
    \kappa \frac{d\langle\overline{c}\rangle}{dy}
    +
    v_s\langle\overline{c}\rangle
    =
    \langle \overline{v'c'} \rangle,
    \label{eq:c mean}%
\end{gather}
with $\langle \overline{u} \rangle = u = y$ and $\langle \overline{c} \rangle =
c = \frac{v_s}{\kappa}\exp[v_s (1 - y) / \kappa] / \sinh(v_s/\kappa)$, when flow
is laminar.  The streamwise sediment fluxes for these solutions are included in
Fig.~\ref{fig:transport}(a) in gray, for the two $v_s$ values corresponding to
the example WS and SBL flows.  In each case, the flux is everywhere enhanced by
turbulence.  In the WS flow, this is primarily due to turbulent vertical
concentration fluxes making more sediment available in the upper channel.
In the SBL flow, this effect is important too---streamwise flux in
the region $y > 0.4$ accounts for roughly one third of the total flux $\frac{1}{2}\int_0^2
\langle \overline{uc} \rangle \,dy$.  Furthermore, via Eq.~\eqref{eq:u mean},
turbulent modification of the mean flow [plotted in Fig.~\ref{fig:profiles}(b)]
by the Reynolds stress $\langle \overline{u'v'}\rangle$ (not shown) enhances
transport within the boundary layer.  To complement this picture, in
Fig.~\ref{fig:transport}(b), we plot $\langle\overline{v'c'}\rangle$ and
$-\kappa d\langle\overline{c}\rangle/dy$.  We see that the WS concentration
profile is primarily dictated by near constant uplift from vertical turbulent
fluxes [balancing the settling term $v_s\langle\overline{c}\rangle$ in
Eq.~\eqref{eq:c mean}] except very close to the walls, where sediment diffusion
takes over.  In the SBL case, there is a clear separation between the
diffusively dominated boundary layer, where the peak streamwise flux occurs, and
the turbulent suspension in the upper channel.

While turbulence is greatly suppressed within the sediment boundary layer,
it may intrude from the upper channel, where TKE is higher.
Both the mixing of streamwise momentum and the mixing of concentration between
these two regions occur in a spatiotemporally intermittent way. This may be
visualized by viewing channel slices at different depths.  In
Figs.~\ref{fig:intermittency}(a)--(d), we plot horizontal cross-sections of the
$u$ field for the example WS~(a,b) and SBL~(c,d) flows at
$y=0.4$ and $y=0.1$. The WS flow is well developed at $y=0.4$, with streaky
near-wall structures at $y=0.1$ that qualitatively resemble their well-studied
counterparts in unstratified shear
flows~\cite{Moin1982,Robinson1991,Jimenez2004}.
In contrast, the SBL turbulence is patchy at $y=0.4$ (the edge of the sediment
boundary layer). At $y=0.1$, the flow is essentially quiescent, except for an
isolated spot of faster fluid advected into the boundary layer from above.
The plotted snapshots are broadly representative of the sustained flow dynamics,
which feature a tension between the transient proliferation of turbulent structures
and their suppression by high concentration gradients.
To measure this phenomenon, we compute the average turbulent fraction at each
height, by computing the intermittency factor $I(y) = \langle
\overline{\chi(\vect{u})}\rangle$, where $\chi$ is an indicator function defined
to be $1$ if the TKE exceeds $10^{-4}$ and zero otherwise. (Other choices for
$\chi$ lead to qualitatively similar conclusions.) We plot $I(y$) in
Fig.~\ref{fig:intermittency}(e).  For both flows, $I(y) \approx 1$ throughout
the channel interior, falling to $I(y) = 0$ at the walls. The length scale over
which this occurs is dictated by the thickness of the viscous boundary layers for
each fluid, except in the case of SBL flow at the bottom wall, for which $I(y)$
transitions comparatively slowly from $0$ to $1$ over $0 \leq y \lesssim 0.4$.

Since the SBL flow is intermittent in the region containing most of the
sediment, this has implications for the transport.  
%
\begin{figure}
    \includegraphics[width=0.98\columnwidth]{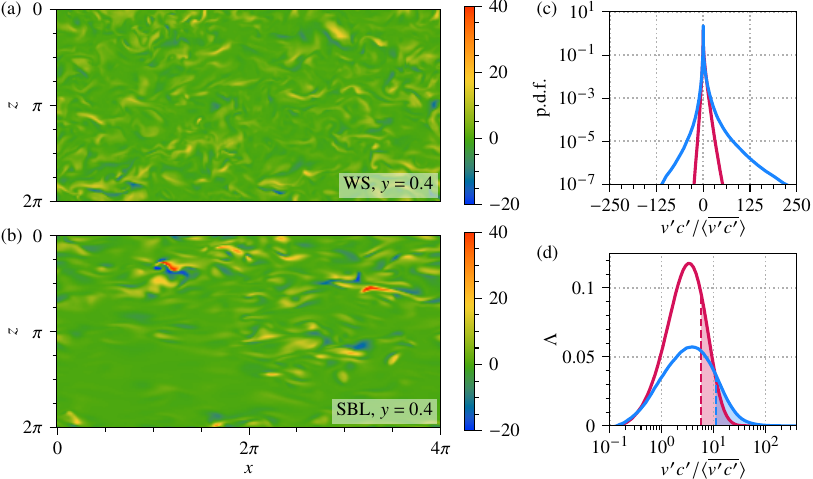}%
    \caption{Intermittency of vertical sediment transport.  (a,b)~Example
    horizontal cross-sections of $v'c'/\langle\overline{v'c'}\rangle$ 
    at $y = 0.4$ for (a)~WS and
    (b)~SBL flows, taken at the same instants as
    Figs.~\ref{fig:intermittency}(a--d).
    The color scale is thresholded between $-20$ and $40$, improving visual
    clarity across the panels. This clips the range of the SBL data,
    which attains a minimum of $-40.8$ and maximum of $86.7$ (3~s.f.).
    Videos of the full simulations are available as supplementary.
    Panels (c) and (d) plot statistics for the WS (solid red) and SBL (solid
    blue) cases:
    (c)~p.d.f of
    $\mu = v'c'/\langle\overline{v'c'}\rangle$;
    (d)~$\Lambda(\mu) = \mu[f(\mu)
- f(-\mu)]$, where $f$ is the p.d.f. of $\mu$.
    The shaded regions divide the area under each curve in half.
    }
    \label{fig:vpcp_dist}%
\end{figure}
In Figs.~\ref{fig:vpcp_dist}(a,b), we compare two illustrative snapshots of
(normalized) vertical flux $v'c'/\langle \overline{v'c'}\rangle$ for (a)~WS and
(b)~SBL flow.  The WS fluxes are comprised of many upward and downward
contributions, which exhibit some streamwise alignment and are distributed
homogeneously throughout the channel.  In contrast, the SBL fluxes consist of a
few comparatively large events scattered within the regions where turbulent
intensity is high [see Fig.~\ref{fig:intermittency}(c)]. In
Fig.~\ref{fig:vpcp_dist}(c), the corresponding probability density function
(p.d.f) is plotted for both flows.  The probability mass of both distributions
is highly concentrated around zero, with rapidly decaying tails that count
rarer, but more significant fluxes. As expected, the tails for SBL flow are
wider and greater magnitude.  In both cases, $\langle \overline{v'c'}\rangle$ is
a sum of nearly canceling positive and negative contributions that are slightly
positively skewed to produce an upward net flux that balances settling and
diffusion [Eq.~\eqref{eq:c mean}].  To identify which events tip the balance, we
plot in Fig~\ref{fig:vpcp_dist}(d) the difference between the positive and
negative halves of the p.d.f., weighted by the corresponding (normalized)
magnitude of vertical flux. This quantifies the net contribution due to upward
fluxes of a given size, to the mean of $v'c'$, after subtracting off parts that
cancel with equal and opposite downward fluxes.  Both datasets are divided into
two groups, which contribute equally to the mean.  Despite their importance to
maintaining the suspension, the higher flux events comprise less than $6\%$ and
$3\%$ of the probability masses of the WS and SBL p.d.f.s respectively.

\begin{figure}
    \includegraphics[width=0.6\columnwidth]{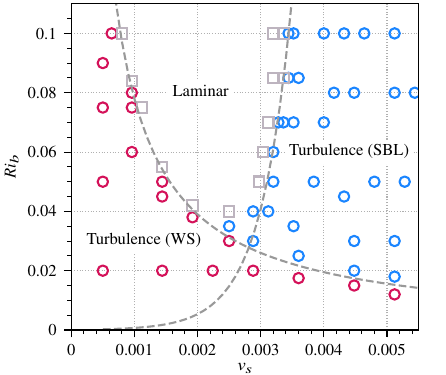}%
   \caption{Laminar--turbulent boundary. Circles plot $(v_s,\Ri_b)$
    pairs, for which DNS maintains steady WS (red) or SBL
    (blue) turbulence.
    Squares show parameters for which we were unable to find
   sustained turbulence. The dashed gray lines 
    are the curves $\Ri_b=A/v_s$ and 
    $\Ri_b=B\exp(v_s \delta / \kappa)$,
    with $A = 7.8\times10^{-5}$, $B = 9.2\times 10^{-5}$ and $\delta=0.65$.}
   \label{fig:lamturbbdry}%
\end{figure}

The observations made thus far qualitatively generalize as $\Ri_b$ and $v_s$ are
varied away from our chosen example values.  To demonstrate this, we have
conducted extensive numerical investigations of parameter space. The data are
summarized in Fig.~\ref{fig:lamturbbdry}.
Points on these axes are plotted with circles if DNS remained in a statistically
steady state with nonzero TKE for at least 1000 advective time units.
Each simulation was initiated from a developed turbulent state at nearby
parameter values (starting from passive scalar flows at $\Ri_b = 0$). Guided by
Fig.~\ref{fig:profiles}(e), we
classify flows as SBL turbulence if $\max_y \Ri_g$ occurs at the lower wall and
WS otherwise. Within each regime, we find flow
statistics that are qualitatively similar to those of the example flows reported
in Figs.~\ref{fig:profiles} and~\ref{fig:transport}.  

Runs that decayed to laminar flow are plotted with squares on
Fig.~\ref{fig:lamturbbdry}. Tracing the laminar-turbulent boundary requires
care, because the process of decay to the laminar state is stochastic. Moreover,
for initial conditions far from the turbulent attractor it can be the case that
turbulence is fully suppressed before the concentration field has time to
statistically equilibrate to a state that would otherwise permit turbulence. For
these reasons, each point straddling the boundary was initiated by varying
$\Ri_b$ by no more than $5\times 10^{-3}$, or $v_s$ by no more than $2\times
10^{-4}$, from an established turbulent flow and integrating for at least 2000
time units. Near the intermittent SBL regime boundary, turbulence can persist
for much longer before decaying, so runs of up to $10\,000$ time units were
employed.

Two fitted curves (dashed gray) plotted on Fig.~\ref{fig:lamturbbdry} separate
the regimes.  In the WS case, turbulence cannot be sustained if $\Ri_b v_s$
exceeds a constant threshold. This scaling was suggested by Cantero \emph{et
al.}~\cite{Cantero2012a} for a similar flow configuration.  On taking the dot
product of Eq.~\eqref{eq:governing 1} with $\vect{u}$ and averaging over space
and time, one obtains an equation for the steady-state turbulent kinetic energy
budget which depends on sediment concentration only via the term $-\Ri_b
\langle\overline{v'c'}\rangle$. This accounts for TKE losses due to vertical
fluxes. The aggregate loss over the channel is obtained by integrating the term
over the wall-direction and using Eq.~\eqref{eq:c mean} to obtain $-\Ri_b [v_s +
\kappa \langle\overline{c}(2) - \overline{c}(0)\rangle] \approx -\Ri_b v_s$.
If this drops too low, turbulence cannot be sustained throughout the channel and
the flow must either fully laminarize, or enough sediment must drop out of
suspension for the turbulence to survive above a partially laminarized region,
as it does in the SBL regime.  In this latter case, turbulence suppression can no
longer be captured by a global balance. The upper channel becomes increasingly
diluted at higher settling velocities, thereby enabling turbulence to
proliferate, and the laminar-turbulent boundary rises acutely. Motivated by the
form of the laminar solution, whose concentration decays exponentially in this
regime, we find that a curve of the form $\Ri_b \propto \exp(v_s \delta /
\kappa)$ (where $\delta$ is determined empirically) separates the data points well.

At higher $\Rey$, it may be anticipated that the laminar region retreats
towards higher $\Ri_b$, just as the corresponding laminarization threshold does
in thermally stratified shear flow~\cite{Deusebio2015}.
However, we hypothesize that the WS and SBL flow regimes qualitatively
persist, since the mechanisms that separate them are not specific to
moderate-$\Rey$ turbulence.
Though the dilute continuum model considered herein does not attempt to
describe the full physics of fluid-sediment interactions, our results
capture some essential features of environmental flows. For example, flows in
river channels have been observed to bifurcate into transport regimes that carry
sediment in suspension and in a localized near-bed layer~\cite{Ma2020}.  In this
setting, the commonly applied diffusive term in Eq.~\eqref{eq:governing 3} could
be viewed as a phenomenological closure, which promotes a basal `reservoir' of
sediment that can be ejected into suspension by turbulent fluctuations.  The
striking spatiotemporal intermittency of this process, as demonstrated in
Fig.~\ref{fig:vpcp_dist}, suggests a need to move beyond descriptions of
sediment transport that average over transient flow structures.

~\\
~\\
\textbf{Supplemental material.}
The supplemental videos show the full time evolution of the simulations from
which the instantaneous data slices in Figs.\ 3(a--d) and 4(a,b) were plotted. To
aid in visualization of the flow structures, the fields have been transformed to
a reference frame that is co-moving with the mean streamwise velocity. (However,
note that the color bar values pertain to the original frame.) Files beginning
with `u' show the streamwise velocity and files beginning with `vpcp' show the
vertical turbulent sediment flux normalized by its mean value (as in Fig. 4).
`L2' and `L3' indicate slices at `$y = 0.4$' and `$y = 0.1$' respectively. Finally,
the suffixes `ws', `sbl' denote weakly self-stratified flow and sediment
boundary layer flow respectively.

~\\
\acknowledgements{
We are grateful to Robert M.\ Dorrell and Charlie J.\ Lloyd for useful early
discussions and for initially prompting us to investigate self-stratifying
flows.  We acknowledge funding from EPSRC New Horizons grant EP/V049054/1.  This
work was carried out using the computational facilities of the Advanced
Computing Research Centre, University of Bristol.
}

%

\end{document}